\documentclass[12pt]{iopart}

\usepackage{iopams}
\usepackage{graphicx}

\begin{document}

\title[Impurity-induced shocks in TASEP with long-range hopping]{Impurity-induced shocks in the asymmetric exclusion process with long-range hopping}

\author{J Szavits-Nossan and K Uzelac}
\address{Institute of Physics, Bijeni\v{c}ka cesta 46, HR-10001 Zagreb, Croatia}
\eads{\mailto{juraj@ifs.hr}, \mailto{katarina@ifs.hr}}

\begin{abstract}
We consider the totally asymmetric simple exclusion process (TASEP) on the periodic chain in the presence of a single impurity site that is inaccessible to other particles and therefore acts as a static defect. Particles are allowed to advance any distance $l\geq 1$ on the right with the probability that decays as $l^{-(1+\sigma)}$, where $\sigma>1$. Despite the long range of hopping, we find the same type of phase transition that occurs in the standard short-range TASEP with a defect site where defect induces a macroscopic shock in the stationary state. In particular, our model displays two main features characteristic of the short-range TASEP with defect site: a growth of the shock width with system size $L$ as $L^{1/2}$ or $L^{1/3}$, depending on the existence of the particle-hole symmetry, and the power-law decay in density profiles of the shock phase. However, unlike the profiles in the short-range case, we find that the latter are well reproduced by the mean-field approximation, which enables us to derive the analytical expression for $\sigma$-dependent exponent $\nu=\sigma-1$ of this power-law decay and the point $\sigma_c=4/3$ at which the transition takes place.
\end{abstract}



\maketitle

\section{Introduction}

Driven non-equilibrium systems have been a subject of extensive study in the past few decades. Among many others, systems of interest include growing interfaces \cite{Krug97}, vehicular traffic \cite{Chowdhury00} and biological transport \cite{Chowdhury05}. One of the simplest of related models is the totally asymmetric simple exclusion process (TASEP) \cite{Derrida98}, in which particles jump to the nearest-neighbour site on the right, while each site holds at most one particle. In case of the open boundary conditions, the finding of the exact stationary state \cite{SchutzDomany93,DerridaEvans93} exhibiting boundary-induced phase transitions \cite{Krug91} has triggered a study of various generalizations including Langmuir kinetics \cite{ParmeggianiFranoschFrey03,ParmeggianiFranoschFrey04} or inhomogeneous hopping rates \cite{Barma06}. The last example is of a particular interest in studying the effects of disorder (see, for example, \cite{Stinchcombe02}), which is hard to avoid in most realistic situations.

Disorder in TASEP is usually introduced through inhomogeneous hopping rates that are either associated to particles or sites. Due to the long-range correlations already present in the pure TASEP, strong effects are found even in the presence of a singe defect particle \cite{Mallick96} or site \cite{JanowskyLebowitz92, Schutz93, HinrichsenSandow97, SlaninaKotrla98, Kolomeisky98, Ha03}. In the case of a single defect particle, the exact solution was found using the matrix-product ansatz \cite{Mallick96} and was recently generalized to arbitrary number of defects \cite{EvansFerrariMallick09}. On the other hand, TASEP with defect sites (and even a single one) is still an open problem, usually dealt with Monte Carlo simulations and various mean-field approaches (see \cite{Foulaadvand08} and references therein). 

There is, however, a particular generalization of TASEP which reproduces its main features, but is remarkably well described by the mean-field approach. Originally motivated by the emergence of long-ranged power-law-type correlations in TASEP, a generalized model was proposed \cite{SzavitsUzelac06} in which particles are allowed to advance any distance $l\geq 1$ with the probability that decays as $l^{-(1+\sigma)}$. In case of the open boundary conditions, albeit modified due to the long range of hopping, this generalized model has the same phase diagram as the short-range case consisting of the low-density, high-density and the maximum-current phase, but with different effects at the transition lines. Careful analysis of this model \cite{SzavitsUzelac08} reveals the emergence of an effective bulk reservoir (similar to the one introduced by Langmuir kinetics), which for $1<\sigma<2$ dominates over the fluctuations.   

Similarity with the short-range TASEP and the applicability of the mean-field approach motivates us to study the effect of a single impurity also in the long-range case. At first sight, one could expect that a localized defect in system with long-range hopping would not have any significant effect on the flow of particles. The main objective of this paper is to show that this is not the case. For that purpose we introduce the simplest defect in terms of a static impurity that occupies one site on the lattice and is not allowed to hold any particle. This induces the same type of transition as in the short-range case, but with respect to $\sigma$ as a control parameter. We show that the model is well described by the mean-field approach which enables us also to obtain the analytical expression for $\sigma_c$ at which the transition takes place.

The paper is organized as follows. In \sref{sec2} we define the exclusion process with long-range hopping on the periodic lattice and introduce a defect site to study its effects on the stationary states. Density profiles and their scaling properties obtained by Monte Carlo simulations are displayed in \sref{sec3mc} and analyzed within the mean-field approach in \sref{sec3mf}. A brief summary of results is given in \sref{conc}.

\section{Exclusion process with long-range hopping and impurity}
\label{sec2}

In the pure long-range model without defect \cite{SzavitsUzelac06}, $N=\rho L$ particles are distributed on $L$ sites of a one-dimensional lattice with periodic boundary conditions such that each site is either occupied by a particle ($\tau_n=1$) or empty ($\tau_n=0$). During an infinitesimal interval $dt$, a randomly chosen particle at site $1\leq n\leq L$ attempts to jump $l$ sites to the right, where distance $1\leq l\leq L$ is chosen according to the probability distribution $p_{l}=l^{-(1+\sigma)}/\zeta_L(\sigma+1)$ with $\zeta_L(z)$ being the partial sum of the Riemann zeta function. If the target site $n+l$ is empty, the move is accepted; otherwise, it is rejected. Compared to the standard (short-range) TASEP, common features include factorized steady state and the particle current of the same form, $j=\lambda(\sigma)\rho(1-\rho)$, where the additional factor $\lambda(\sigma)=\sum_l l\cdot p_{l}=\zeta_{L}(\sigma)/\zeta_{L}(\sigma+1)$ may be interpreted as the average hopping length. The model is well defined for $\sigma>1$ where $\lambda(\sigma)$ is finite. For $1<\sigma<2$ it displays true long-range behaviour, while the short-range regime sets in for $\sigma>2$ \cite{SzavitsUzelac06, SzavitsUzelac08}.

To study the simplest case in which the flow of particles is obstructed by a single defect, an additional site is included at the position $L+1$ and the periodic boundary conditions are shifted accordingly ($\tau_n=\tau_{n+L+1}$). The additional site is not allowed to hold any particle, but the particles are still able to pass it as long as $\sigma<\infty$ (the flow is fully blocked only in the short-range limit $\sigma\rightarrow\infty$ where $p_l=\delta_{l,1}$). The key benefit from such definition of impurity is that it obstructs the flow of particles but does not introduce any new parameter in the model. Any change in behaviour therefore must come solely from the variation of $\sigma$.

\section{Density profiles}\label{sec3}
\subsection{Monte Carlo simulations}\label{sec3mc}

\begin{figure}[!ht]
\centering\includegraphics{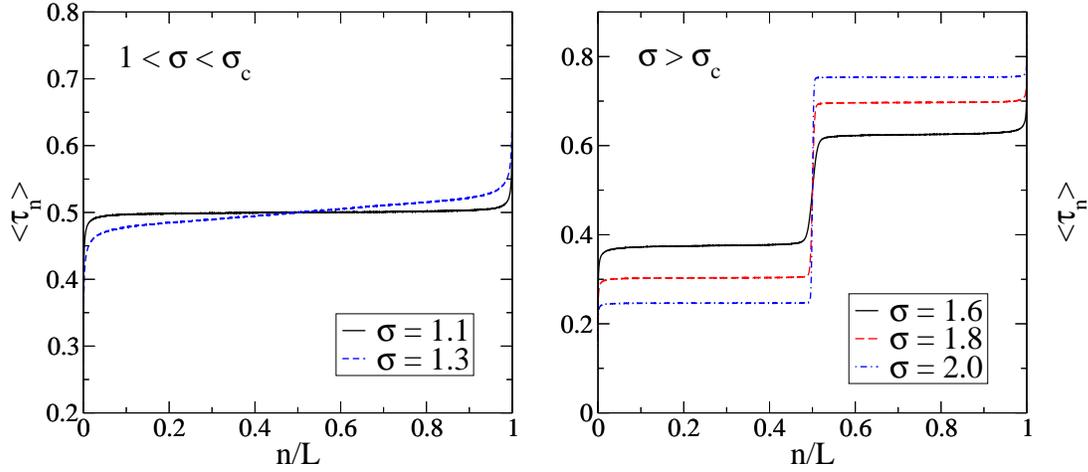}
\caption{Density profiles in the (a) shock-free phase for $\sigma=1.1$ and (b) shock phase for $\sigma=1.8$ obtained by Monte Carlo simulations for $L=6400$, $\rho=1/2$ and $t=10^8$ MCS/site.}
\label{fig1}
\end{figure}

Typical density profiles obtained by Monte Carlo simulations for $\rho=1/2$ are shown in \fref{fig1}. As in the short-range case, one observes two regimes (separated by the value $\sigma_c$ of the control parameter): for $\sigma>\sigma_c$ the defect is strong enough to induce shocks with densities $\rho_+$ and $\rho_-=1-\rho_+$ left and right of the impurity, respectively; for $1<\sigma<\sigma_c$ the density profile is shock-free. In the latter case the profile organizes itself in a way that maximizes the current, which attains the maximal value as in the pure long-range model, $\lambda(\sigma)/4$.

%
%

For any finite system the shock is also characterized by its width, as a result of the fluctuations in the microscopic position of the instantaneous shock that performs random walk around the lattice. Using the useful concept of the second-class particle \cite{JanowskyLebowitz92} it was found that the width in the short-range model scales with $L$ as $L^{1/2}$ and $L^{1/3}$ depending on whether $\rho\neq 1/2$ or $\rho=1/2$, respectively, which is related to whether the collective velocity $v_g=1-2\rho$ at which the density fluctuations travel is finite or vanishes \cite{JanowskyLebowitz92}. 

\begin{figure}[!ht]
\centering\includegraphics{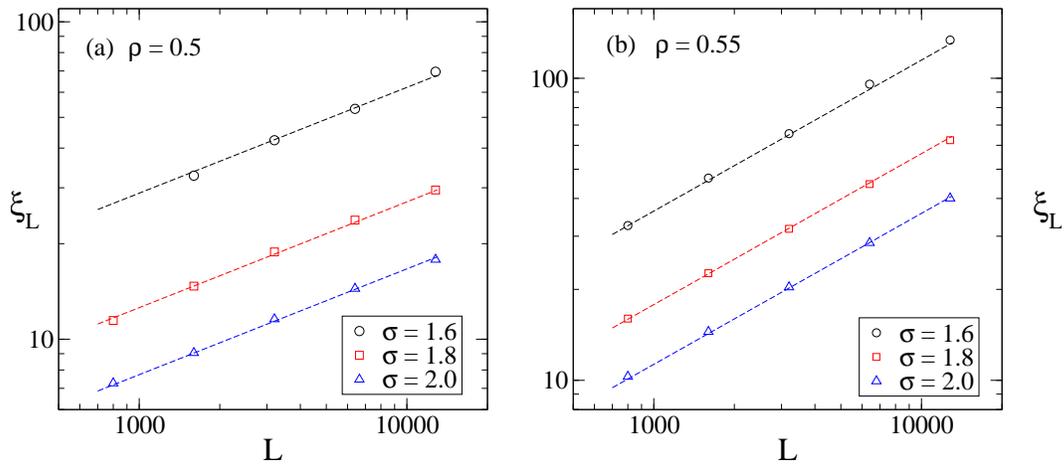}
\caption{Shock width $\xi_L$ obtained from the stationary density profile for various $\sigma$ and densities (a) $\rho=0.5$ and (b) $\rho=0.55$. Dashed lines have slopes (a) $1/3$ and (b) $1/2$.}
\label{fig2}
\end{figure}

We recover this result also in the long-range model, in accordance with our earlier result \cite{SzavitsUzelac08} that the density fluctuations in the infinite system travel with the velocity $v_g=\lambda(\sigma)(1-2\rho)$. In \fref{fig2} we plot the width $\xi_L$, obtained by the Gaussian fit of the discrete derivative $\langle\tau_{n+1}\rangle-\langle\tau_{n}\rangle$ vs. system size $L$ for densities $\rho=0.5$ (\fref{fig2}a) and $0.55$ (\fref{fig2}b). 

%
%

Another interesting feature to compare with the short-range model are the site-dependent corrections to the bulk densities $\rho_-$ and $\rho_+$. In the short-range model it was found that they exhibit power-law behaviour away from the defect site

\begin{equation}
	\label{profile_sr}
	\langle\tau_n\rangle-\rho_{-}\sim -n^{-\nu}, \qquad \langle\tau_{L-n}\rangle-\rho_{+}\sim n^{-\nu}, \quad 1\ll n\ll L,
\end{equation}

\noindent while different values of $\nu$ were given in literature. Early results of Janowsky and Lebowitz for the periodic boundary conditions proposed that $\nu=1$ \cite{JanowskyLebowitz92} based on the Monte Carlo simulations. The problem was later studied in the context of surface growth \cite{SlaninaKotrla98} in the particular model that can be mapped directly to TASEP with one defect site. They considered only the shock-free phase and obtained the exponent $\nu=0.41$ using renormalization-group-like treatment of the exact solution for $N=2$ and the Monte Carlo simulations. These results were later revised by Ha et al. \cite{Ha03} based on numerical results and analytical arguments arguing that the power-law behaviour (\ref{profile_sr}) occurs in both phases, with $\nu=1/2$ in the shock phase ($\rho_{-}\leq\rho_{+}$) and $\nu=1/3$ in the shock-free phase ($\rho_{-}=\rho_{+}=1/2$) where the current is maximal.

In order to check the power-law behaviour (\ref{profile_sr}) in our model, we assume the following scaling relation for a deviation of density from its bulk value $\langle\tau_{L/4}\rangle$ 

\begin{equation}
\label{profile_lr}
\Delta\rho_{-}(n,L)\equiv\langle\tau_n\rangle-\langle\tau_{L/4}\rangle=L^{-\nu}f_{-}(n/L).
\end{equation}

%
%

\begin{figure}[!ht]
\centering\includegraphics{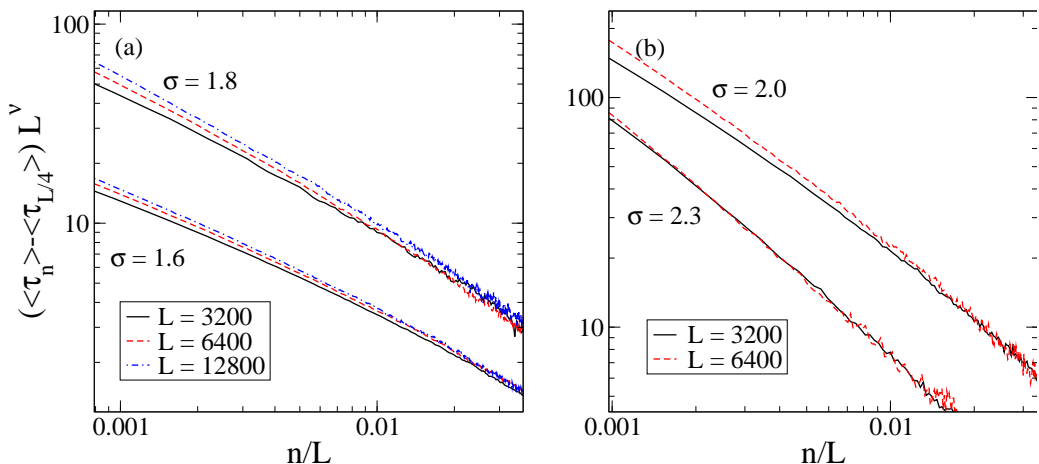}
\caption{Deviation of a density profile from its bulk value $\langle\tau_{L/4}\rangle$ in the low-density domain of the shock phase for (a) $\sigma=1.6$ and $1.8$ and (b) $\sigma=2.0$ and $2.3$. Profiles have been scaled for different system sizes $3200$, $6400$ and $12800$ according to (\ref{profile_lr}) with (a) $\nu=0.6$ and $0.8$ and (b) $\nu=1$, respectively ($\rho=1/2$ and $t=10^8$ MCS/site).}
\label{fig3}
\end{figure} 

\noindent Scaled density profiles are shown in \fref{fig3}a for $\sigma=1.6$ and $1.8$ and various system sizes. The best collapsing fit is achieved for $\nu\approx 0.6$ and $\nu\approx 0.8$ respectively, suggesting that $\nu$ is equal to $\sigma-1$. On the other hand, for $\sigma>2$ one generally expects the short-range regime to set in \cite{SzavitsUzelac08} with a $\sigma$-independent $\nu$. Indeed, \fref{fig3}b for $\sigma=2$ and $\sigma=2.3$ shows good agreement between the scaled profiles if one assumes that $\nu(\sigma)=\nu(2)=1$. This exponent matches the one suggested by Lebowitz and Janowsky in the short-range TASEP \cite{JanowskyLebowitz92}. Let us also note here that in the shock-free phase the scaling is not so clear and remains elusive. Although we observe density profiles that certainly exhibit long-range correlations, they are not described by the single exponent $\nu$ as in (\ref{profile_lr}). We shall return to this point later in the discussion of the mean-field approach.

\begin{figure}[!ht]
\centering\includegraphics{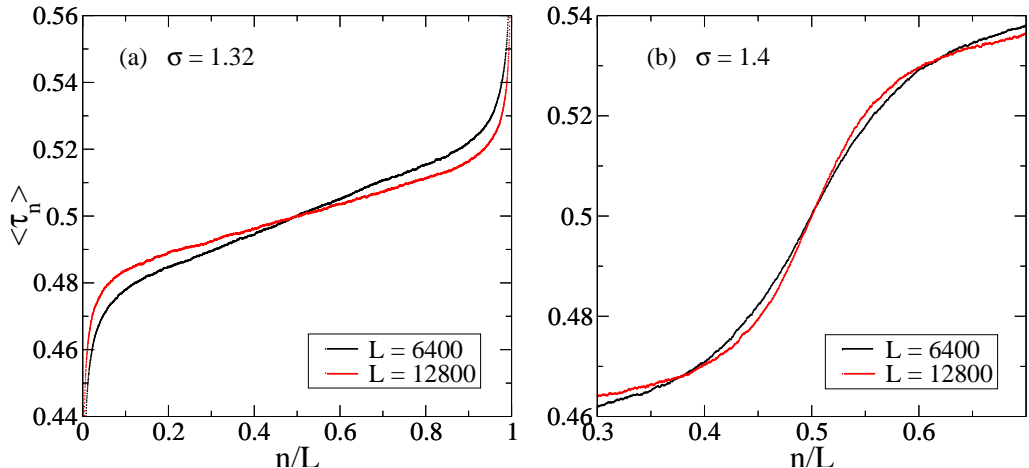}
\caption{Emergence of the shock-free phase (a) for $\sigma=1.32$ and shock phase (b) for $\sigma=1.4$ from the density profiles on large systems, $L=6400$ and $12800$ ($\rho=1/2$ and $t=10^8$ MCS/site.}
\label{fig4}
\end{figure}

Finally, let us touch upon the value of $\sigma_c$. The main problem in finding $\sigma_c$ numerically are the finite-size effects that may create shock for any $\sigma$ provided that the system size is small enough. Near $\sigma_c$, system sizes needed to observe the transition become inaccessible to computer simulations and one has to relay on other methods. For example, in the short-range case where the transition takes place at value $r_c$, Janowsky and Lebowitz suggested the estimate $r_c\gtrsim 0.8$ based on the exact expression for the current $j_L(r)$ obtained for small systems \cite{JanowskyLebowitz94}. On the other hand, Ha et al. \cite{Ha03} assumed the finite-size scaling of the order parameter $\rho_{+}(r,L)-\rho_{-}(r,L)$ and obtained the best collapse of data for $r_c=0.80(1)$. Here we take a different and less precise approach by looking at the smallest $\sigma$ for which the slope of the shock in the middle becomes steeper as one increases the system size $L$. In other words, for $\sigma<\sigma_c$ increasing $L$ will flatten up the shock, while for $\sigma>\sigma_c$ one expects convergence to the step-like function. For the largest system size that we used in Monte Carlo simulations, $L=12800$, this gives the estimate $1.32<\sigma_c<1.4$ (\fref{fig4}). 

%
%

\subsection{Mean-field approach}\label{sec3mf}

The mean-field approximation, in which one neglects correlations by replacing $\langle\tau_n\tau_m\rangle$ with $\langle\tau_n\rangle\langle\tau_m\rangle$, has been the starting point for many of the analytical approaches to the problem of defect site in TASEP (with random-sequential update) so far. The reason for this lies in the inapplicability of the usual tools like the matrix-product ansatz that proved to be extremely useful in the pure TASEP. Here we take the same approach of neglecting correlations between sites, which unlike the short-range TASEP gave accurate results in our earlier study of the pure long-range case \cite{SzavitsUzelac08}.

Before writing down the lattice equations for density profile $\langle\tau_n\rangle$ and applying the usual mean-field approximation $\langle\tau_n\tau_m\rangle\rightarrow\langle\tau_n\rangle\langle\tau_m\rangle$, let us recall the simple mean-field (SMF) approximation dealing only with bulk values \cite{JanowskyLebowitz92} to estimate $\rho_-$ and $\rho_+$. In particular, we neglect any correlations between sites and assume the following density profile in the shock phase

\begin{equation}
\label{smf}
\langle\tau_n\rangle=\cases{\rho_{-},&$1\leq n\leq L/2$\\ \rho_{+},& $L/2+1\leq n\leq L$.}
\end{equation}

\noindent We then use the conservation of current to compare the current at two different sites, one at the impurity  and the other far in the bulk of the low- or high-density domain. In the short-range model this gives densities $\rho_-=r/(1+r)$ and $\rho_+=1/(1+r)$, where $0 < r\leq 1$ is the reduced hopping rate characterizing the defect there. In the long-range case, the conserved quantity is the site-independent particle current defined as a summation over all the particles jumping over and through some site $k$

\begin{equation}
	j=\sum_{n=1}^{k}\sum_{l=k+1-n}^{L}p_l\langle\tau_n(1-\tau_{n+l})\rangle+\sum_{n=k+2}^{L}\sum_{l=L-n+k+2}^{L}p_l\langle\tau_n(1-\tau_{n+l})\rangle.
\end{equation}

\noindent  In the bulk of the low-density domain, i.e. far away from the impurity ($k=L/4$), the current $j$ is roughly $\lambda_L(\sigma)\rho_{-}(1-\rho_{-})$. On the other hand, the current over the impurity (obtained by setting $k=L$) is given approximately by $(\lambda_{L}(\sigma)-1)\rho_{+}(1-\rho_{-})$. [In both cases we have neglected terms that arise from the jumps longer than $L/2$, but they vanish in the limit $L\rightarrow\infty$ anyway.] Equating these two expressions gives $\rho_-$ and $\rho_+$ for a very large system in terms of $\lambda=\sum_{l=1}^{\infty}l p_l$

\begin{equation}
	\label{rho_smf}
	\rho_-=\frac{1-\lambda^{-1}}{2-\lambda^{-1}}, \qquad \rho_+=\frac{1}{2-\lambda^{-1}},
\end{equation}

\noindent Equation (\ref{rho_smf}) may be reduced to the form of the short-range model, $\rho_-=r'/(r'+1)$ and $\rho_+=1/(r'+1)$, where the reduced hopping rate $r'$ is given by $r'=(\lambda-1)/\lambda$. One way to interpret this expression is as the ratio of the average hopping length over the defect site (effectively reduced by one due to the defect site), $\sum_l (l-1) p_l$, versus the average hopping length in the absence (or far from) defect, $\sum_l l p_l$. 

\begin{figure}[!ht]
\centering\includegraphics{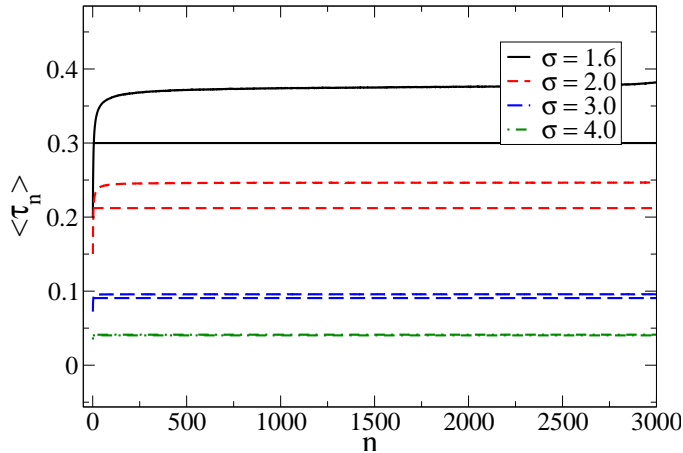}
\caption{A comparison of bulk densities $\rho_-$ in the shock phase obtained by the simple mean-field approach (horizontal lines) and by Monte Carlo simulations ($L=6400$, $\rho=1/2$ and $t=10^8$ MCS/site).}
\label{fig5}
\end{figure}

Compared to the results of Monte Carlo simulations, we find that the above result becomes accurate only for large values of $\sigma$, $\sigma\gtrsim 3$ (\fref{fig5}). Therefore, it is not surprising that equating $\rho_{-}(\sigma_c)=\rho_{+}(\sigma_c)$ in the above expression to obtain $\sigma_c$ gives trivial value $\sigma_{c}^{SMF}=1$ ruling out the shock-free phase completely. Such a crude approximation that does not allow spatial variation of density cannot produce non-trivial transition to the shock-free phase in the short-range TASEP as well, where it gives $r_{c}^{SMF}=1$ \cite{JanowskyLebowitz92}.

%
%

\begin{figure}[!ht]
\centering\includegraphics{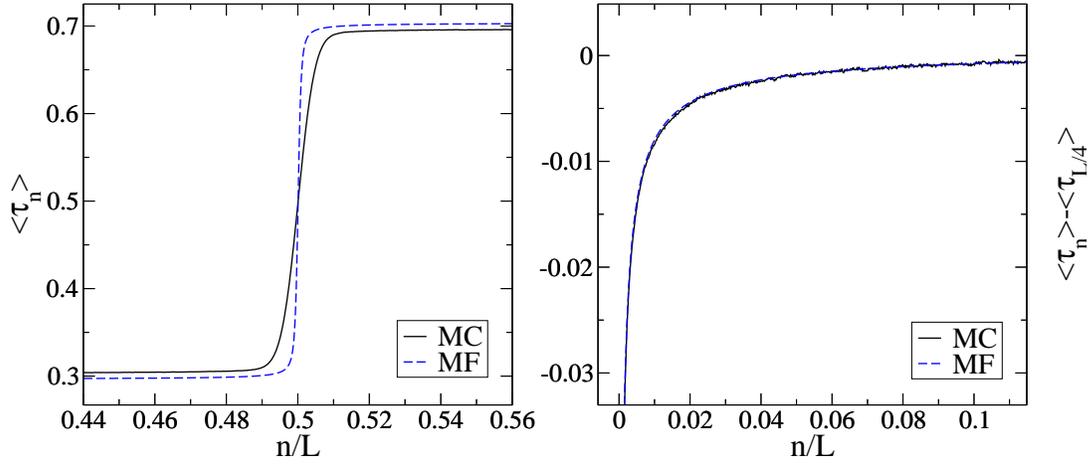}
\caption{(a) A comparison of density profiles obtained in the mean-field approximation and by Monte Carlo simulations for $\sigma=1.8$ 
($L=6400$, $\rho=1/2$ and $t=10^8$ MCS/site). (b) The site-dependent corrections to the bulk density $\langle\tau_{L/4}\rangle$ obtained in the mean-field approximation coincide with the results of the Monte Carlo simulations.}
\label{fig6}
\end{figure}

\begin{figure}[!ht]
\centering\includegraphics{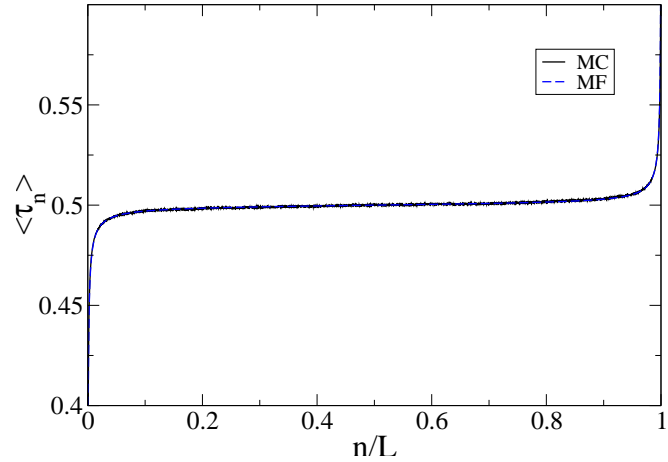}
\caption{(a) A comparison of density profiles obtained in the mean-field approximation and by Monte Carlo simulations for $\sigma=1.1$ 
($L=6400$, $\rho=1/2$ and $t=10^8$ MCS/site).}
\label{fig7}
\end{figure}

In order to explain the nontrivial value of $\sigma_c>1$ and to study the scaling properties of the profile, we have to consider the site-dependent version of the mean-field approach. Let us first renumber the lattice sites so that $n=-K,\dots,K$, $L=2K$ with an impurity at the site $n=0$. Starting from the master equation, lattice equations for the time-dependent average density profile $\langle\tau_n\rangle(t)$ read

\begin{equation}
\label{tau_n}
\frac{d}{dt}\langle\tau_n\rangle=\sum_{l=1 \atop l\neq n}^{L} p_l\langle\tau_{n-l}(1-\tau_n)\rangle-\sum_{l=1 \atop l\neq L-n+1}^{L} p_l\langle\tau_n(1-\tau_{n+l})\rangle,
\end{equation}

\noindent where periodic boundary conditions have been assumed, $\tau_{K+n}=\tau_{-K+n-1}$. In the mean-field approximation where $\langle\tau_n\tau_m\rangle\rightarrow\langle\tau_n\rangle\langle\tau_m\rangle$, the lattice equations (\ref{tau_n}) in the stationary limit $d\langle\tau_n\rangle/dt\rightarrow 0$ reduce to  a system of $L$ nonlinear equations in $L$ variables restricted to the condition $\sum_{n=1}^{L}\langle\tau_{n}\rangle=N$, which can be solved numerically. For this purpose, we used the HYBRD algorithm taken from the MINPAC library \footnote{http://www.netlib.org/minpack/} for various $\sigma$ and system sizes $L$. In the shock phase, the results reproduce those of the Monte Carlo simulations, but one generally finds a narrower shock and a shift in the density profile in each of the domains (\fref{fig6}a). The first result is not surprising, as the fluctuations in the position of the microscopic shock are usually neglected in the mean-field approximation. Indeed, the width of the shock obtained from the mean-field data seems to obey the $L^{1-\sigma/2}$ scaling compared to the $L^{1/3}$ of the Monte Carlo data. On the other hand, we have no explanation for the shift at this moment, but we note that it is observed in the short-range TASEP with a defect site as well \cite{JanowskyLebowitz92}. Nevertheless, we find the site-dependent corrections in the region dominated by the long-range correlations described correctly in the mean-field approximation (\fref{fig6}b). This becomes more apparent in the shock-free phase, where $\rho_-$ and $\rho_+$ become $1/2$ and the profiles coincide with the results of the Monte Carlo simulations (\fref{fig7}).
%
%

%
%

In consequence, the same scaling of the profile decay from the impurity observed in MC simulations is reproduced in the mean-field approximation as well. Similar behaviour was observed in the pure case of this model with open boundary conditions \cite{SzavitsUzelac08}, and indicates that the long-range hopping has suppressed the importance of fluctuations to the point that the mean-field approximation produces the correct scaling exponent. This motivates us to analyze the leading power-law contributions in the limit of large $L$, and obtain the analytic expression for the exponent $\nu$. To this end we define the site-dependent corrections $\phi_n$ to the (unknown) bulk densities $\rho_{-}$ and $\rho_{+}$ in the stationary regime ($d\langle\tau_n\rangle/dt\rightarrow 0$)

\begin{equation}
\label{phi_n}
\langle\tau_n\rangle=\cases{\rho_{+}+\phi_{n},&$-K\leq n<0$\\ \rho_{-}+\phi_{n},&$0<n\leq K$.}
\end{equation}

\noindent For $0<n\leq K$, the r.h.s. of (\ref{tau_n}) gives the following terms in powers of $\phi_n$

\begin{eqnarray}\label{phi_012}
\fl \phi_{n}^{0}:\quad-\rho_{-}(1-\rho_{-})(p_n-p_{L-n+1})+(\rho_{+}-\rho_{-})\left[(1-\rho_{-})\sum_{l=n+1}^{n+K}p_l+\rho_{-}\sum_{K-n+1}^{L-n}p_l\right]{}\nonumber\\
\fl \phi_{n}^{1}:\quad\left[\rho_{-}\sum_{l=1 \atop l\neq L-n+1}^{L} p_l\Delta_{l}^{+}\phi_n-(1-\rho_{-})\sum_{l=1 \atop l\neq n}^{L}p_l\Delta_{l}^{-}\phi_n\right]-\phi_n(1-2\rho_{-})(p_n-p_{L-n+1})+{}\nonumber\\
{}+(\rho_{+}-\rho_{-})\phi_n\left[\sum_{l=K-n+1}^{L-n}p_l-\sum_{l=n+1}^{K+n}p_l\right]{}\nonumber\\
\fl{}\phi_{n}\phi_{m}:\quad\phi_{n}^{2}(p_n-p_{L-n+1})+\phi_n\left[\sum_{l=1 \atop l\neq L-n+1}^{L}p_l\Delta_{l}^{+}\phi_n+\sum_{l=1 \atop l\neq n}^{L}p_l\Delta_{l}^{-}\phi_n\right],
\end{eqnarray}

\noindent where the following notation has been used, $\Delta_l^{+}\phi_n\equiv\phi_{n+l}-\phi_n$ and $\Delta_l^{-}\phi_n\equiv\phi_{n}-\phi_{n-l}$. Although these equations cannot be solved analytically, one can estimate the relevance of each term by taking the lattice spacing $a=1/L$ as a small parameter and comparing the order of each term in the limit $a\rightarrow 0$, $na\rightarrow x<\infty$. First term in $\phi_{n}^{0}$ is of the order $\Or(a^{\sigma+1})$, because $p_n\sim L^{-(\sigma+1)}x^{-(\sigma+1)}$. For $0<n<L/4$, the second term in $\phi_{n}^{0}$ is positive and of the order $\Or(a^{\sigma})$. To estimate the $\phi_{n}^{1}$ terms, we assume the scaling form $\phi_n=L^{-\nu}f_{-}(n/L)$ with an unknown exponent $\nu$ and $f_{-}(n/L)<0$ for $0<n<L/4$. The second and the third terms in $\phi_{n}^{1}$ are then of the order $\Or(a^{\nu+\sigma+1})$ and $-\Or(a^{\nu+\sigma})$, respectively. Neglecting the higher-order (nonlinear) terms $\phi_n\phi_m$, the only nontrivial terms to estimate are

\begin{equation}
\fl\qquad\frac{1}{2}\left[\sum_{l=1 \atop l\neq L-n+1}^{L}p_l\Delta_{l}^{+}\phi_n-\sum_{l=1 \atop l\neq n}^{L}p_l\Delta_{l}^{-}\phi_n\right]-
\Delta\rho_{-}\left[\sum_{l=1 \atop l\neq L-n+1}^{L}p_l\Delta_{l}^{+}\phi_n+\sum_{l=1 \atop l\neq n}^{L}p_l\Delta_{l}^{-}\phi_n\right],
\end{equation} 

\noindent where we introduced $\Delta\rho_{-}=1/2-\rho_{-}$ to distinguish two cases, $\rho_{-}=1/2$ ($\Delta\rho_{-}=0$) and $\rho_{-}\neq 1/2$ ($\Delta\rho_{-}\neq 0$). In the pure long-range hopping on the infinite lattice with density of particles $\rho$, these terms can be estimated in the continuous limit $a\rightarrow 0$, $x=na$, $\phi_n\rightarrow\phi(x)$ \cite{SzavitsUzelac08}

\begin{equation}
\label{diffusion}
\fl\qquad\frac{1}{2}\left[\sum_{l>0}p_l\Delta_{l}^{+}\phi_n-\sum_{l>0}p_l\Delta_{l}^{-}\phi_n\right]\rightarrow \cases{a^{\sigma}D_{\sigma}\Delta_{\sigma}\phi(x)+\Or(a^2),& $1<\sigma<2$\\ a^2 D_2\Delta\phi(x)+\Or(a^{min\{\sigma,3\}}),& $\sigma>2$,}
\end{equation}

\begin{equation}
\label{drift}
\fl\qquad\Delta\rho\left[\sum_{l>0}p_l\Delta_{l}^{+}\phi_n+\sum_{l>0}p_l\Delta_{l}^{-}\phi_n\right]\rightarrow -a(1-2\rho)\lambda(\sigma)\frac{\partial\phi}{\partial x}+\Or(a^\sigma),
\end{equation}

\noindent where $D_{\sigma}=-\Gamma(-\sigma)cos(\pi\sigma/2)/\zeta(\sigma+1)$, $D_2=\zeta(\sigma-1)/(2\zeta(\sigma+1))$ and $\Delta_{\sigma}$ denotes fractional Laplacian (also referred to as the Riesz fractional derivative \cite{SamkoKilbasMarichev93}) defined with respect to the Fourier transform $\mathcal{F}\lbrace\Delta_{\sigma}f(x)\rbrace=-\vert k\vert^{\sigma}\hat{f}(k)$, $\hat{f}(k)=\mathcal{F}\lbrace f(x)\rbrace$. For $1<\sigma<2$, the leading term in (\ref{diffusion}) refers to the space-fractional diffusion $\Delta_{\sigma}\phi$, typically found in the anomalous diffusion mediated by Levy flights (for a recent review see \cite{MetzlerKlafter04}). On the other hand, the leading contribution to the drift term (\ref{drift}) is local for all $\sigma>1$, but the $\sigma$-dependence is present in the collective velocity $(1-2\rho)\lambda(\sigma)$. 

Although the presence of impurity breaks the translational invariance and the continuous limit no longer holds, we expect that the estimation of the leading terms in (\ref{diffusion}) and (\ref{drift}) is still correct and given by $\Or(a^{\sigma+\nu})$ and -$\Or(a^{\nu+1})$, respectively. Finally, turning back to the r.h.s. of (\ref{phi_n}) and taking the two lowest-order terms with opposite sign, the second term in $\phi_{n}^{0}$ and the first term in (\ref{drift}), we obtain the exponent $\nu=\sigma-1$ for all $\sigma$ for which $\rho_{-}\neq \rho_{+}$ (i.e for $\sigma>\sigma_c$). For $\sigma_c<\sigma<2$, this result is exactly the same as the one conjectured from the Monte Carlo simulations, but it breaks down for $\sigma>2$ where the correlations become important and $\nu$ takes the constant value $\nu=1$.

The shock-free phase is obtained by inserting $\rho_{-}=\rho_{+}=1/2$ in (\ref{phi_012}), which gives only four terms

\begin{eqnarray}\label{phi_012_0.5}
0&=&\frac{1}{4}(p_n-p_{L-n+1})+\frac{1}{2}\left[\sum_{l=1 \atop l\neq L-n+1}^{L}p_l\Delta_{l}^{+}\phi_n-\sum_{l=1 \atop l\neq n}^{L}p_l\Delta_{l}^{-}\phi_n\right]+{}\nonumber\\
&&+{}\phi_{n}^{2}(p_n-p_{L-n+1})+\phi_n\left[\sum_{l=1 \atop l\neq L-n+1}^{L}p_l\Delta_{l}^{+}\phi_n+\sum_{l=1 \atop l\neq n}^{L}p_l\Delta_{l}^{-}\phi_n\right].
\end{eqnarray}

\noindent Since the drift term vanishes, the nonlinear effects become important and the density profile may have more complex form than the one given by the power-law behaviour with a single exponent $\nu$. This is indeed observed in the results of the Monte Carlo simulations, for which the assumed scaling relation $\phi_n=L^{-\nu}f(n/L)$ no longer holds.

So far we established that the above mean-field analysis predicts the leading contribution to the exponent $\nu(\sigma)$ as long as $\sigma_c<\sigma<2$, but it gives no analytical expression for the bulk densities $\rho_{-}$ and $\rho_{+}$ which would enable us to calculate the transition point $\sigma_c$. To this end we use a different argument which relies upon the scaling properties. If we suppose that there is no discontinuity in the exponent $\nu(\sigma)$ at the transition point $\sigma_c$ and use the result, argued below, that $\nu = 1/3$ at the onset of the shock-free regime, $\sigma_c$ is obtained by equating $\nu(\sigma_c) = \sigma_c-1 = 1/3$. This gives $\sigma_c=4/3$, which is also within the bounds $1.32<\sigma_c<1.4$ estimated from the Monte Carlo simulations on large system sizes.

Basically, the argument for $\nu(\sigma_c)=1/3$ is the same as the one that gives $\nu=1/3$ in the shock-free phase of the short-range model \cite{Ha03}. First, one notices that the processing of density fluctuations across the impurity is reduced only for positive fluctuations, leading to the excess of particles in front and depletion of particles behind the impurity. If the centre of mass of density fluctuations travels faster than it spreads, one can ignore the spreading and estimate the number of excess particles solely by looking at the number of such fluctuations present in the system at some given moment. This number, however, cannot exceed the time $t_f$ needed for the fluctuation to travel across the system. Since on average the positive and the negative fluctuations cancel out, the number of excess particles $\delta N$ scales with $t_f$ as $\delta N\sim t_f^{1/2}$. Therefore, we can estimate the number of excess particles by estimating $t_f$. In the shock-free phase, the centre-of-mass velocity of density fluctuations $v_g=\delta j/\delta\rho\approx\lambda[1-2\rho(x)]$ is zero in the bulk, but remains nonzero near the impurity, $v_{g}\sim x^{-\nu}$. The centre-of-mass position thus depends on $t$ as $x_{CM}\sim t^{\frac{1}{\nu+1}}$ and so the time $t_f$ for the fluctuation to travel across the system scales with $L$ as $t_f\sim L^{\nu+1}$. The excess of particles then scales with $L$ as $\delta N\sim t_{f}^{1/2}\sim L^{(\nu+1)/2}$ and the self-consistency implies that $\delta N\sim \int dx x^{-\nu}\sim L^{1-\nu}$, which gives $\nu(\sigma_c)=1/3$. This result is valid only if the spreading of density fluctuations $\xi\sim t^{1/z}$ is less than $x_{CM}\sim t^{\frac{1}{1+\nu}}$, i.e. if $\nu<z-1$, where $z$ is the dynamical exponent. In the long-range case, the exponent $z$ takes the value $z=min\lbrace\sigma,3/2\rbrace$ \cite{Katzav03}, which means that $\sigma_c$ is also the limiting value at which the spreading becomes relevant. [For $\sigma>\sigma_c$ the spreading is irrelevant since the collective velocity $v_g$ is constant and nonzero in the bulk.] 

\section{Conclusion}\label{conc}

In this work we study the asymmetric exclusion process with long-range hopping on the periodic lattice with single impurity site inaccessible to other particles and acting as a defect site. The key advantage of such impurity is that it obstructs the flow of particles but does not introduce any new parameter besides $\sigma$. In spite of the long range of hopping, we show that for $\sigma > \sigma_c$ the impurity induces a macroscopic shock in the stationary state, similarly as in the standard TASEP model with a defect site. In particular, our model reproduces two main features of the short-range TASEP with defect site: namely, (a) the shock width scales with the system size $L$ as $L^{1/2}$ or $L^{1/3}$ depending on the value of density, $\rho\neq 1/2$ or $\rho=1/2$ respectively, and (b) the density profile in the shock phase ($\sigma > \sigma_c$) displays a power-law decay away from the impurity. As in the pure long-range case without a defect, we show that the latter is well reproduced by the mean-field approximation and within this approximation we obtain the analytical expression for the $\sigma$-dependent exponent of the power-law decay 
$\nu=\sigma-1$.
 
Finally, by extending to the present model the argument of \cite{Ha03} that gives $\nu=1/3$ in the shock-free phase of the standard TASEP with a defect site and assuming the continuity of the exponent $\nu(\sigma)$ at the transition point $\sigma_c$, we conjecture on the possible exact value of $\sigma_c=4/3$ and find it to be in a good agreement with our numerical simulations.

Although the model is not in the one-to-one correspondence with the standard TASEP with a defect site but rather shares some of its features, we believe that a clear example of the phase transition taking place at the nontrivial value of the control parameter, as in the presented case, may be relevant to the recent discussions on whether the static blockage in one-dimensional driven diffusive systems always induces a macroscopic shock \cite{Ha03,LeeKim09}.

\ack
This work was supported by the Croatian Ministry of Science, Education and Sports through grant No. 035-0000000-3187. J.S.N. wishes to thank Meesoon Ha for providing him her Ph.D. thesis.

\end{document}